\documentclass[11pt]{article}

\usepackage{moriond,epsfig}

\def\be{\begin{equation}}
\def\ee{\end{equation}}
\def\bea{\begin{eqnarray}}
\def\eea{\end{eqnarray}}

\def\ov{\overline}
\begin{document}
\vspace*{4cm}
\title{$F$-TERM UPLIFTED RACETRACK INFLATION}

\author{ MARCIN BADZIAK }

\address{Institute of Theoretical Physics,
University of Warsaw\\
ul.\ Ho\.za 69, PL--00--681 Warsaw, Poland}

\maketitle\abstracts{
It is shown that two classes of racetrack inflation
models, saddle point and inflection point ones, can be constructed
in a fully supersymmetric framework with the matter field $F$-term
as a source of supersymmetry (SUSY) breaking and uplifting.
Two models of $F$-term SUSY breaking are considered:
the Polonyi model and the quantum corrected O'Raifeartaigh model.
In the former case, both classes of racetrack inflation models
differ significantly from the corresponding models with non-SUSY
uplifting. The main difference is a quite strong dominance of
the inflaton by the matter field. In addition, fine-tuning of the
parameters is relaxed as compared to the original racetrack models.
In the case of the racetrack inflation models coupled to the
O'Raifeartaigh model, the matter field is approximately decoupled from
the inflationary dynamics.}

\section{Introduction}

Recent progress in moduli stabilization, due to the proposal of the KKLT mechanism \cite{kklt}, allowed for constructing viable inflationary models within the string theory. Particularly interesting are those in which the volume modulus drives inflation. In this kind of models, called racetrack inflation, superpotential consists of two non-perturbative terms (originating e.g. from the gaugino condensation in the hidden sector) and a constant contribution from fluxes:
\begin{equation}
\label{Wrace}
W=W_0+Ce^{-cT}+De^{-dT} \,,
\end{equation}
With the use of the above superpotential and the tree-level K\"ahler potential,
\begin{equation}
\label{kahler}
K=-3\ln(T+\overline{T}) \,,
\end{equation}
two different inflationary scenarios have been realized. In the first of them, inflation takes place in the vicinity of a
saddle point of the potential with the axion $\tau$, associated with the
volume modulus, being the inflaton \cite{racetrack}. In the second
scenario, the real part of the volume modulus $t$ is the inflaton and
inflation takes place in the vicinity of an inflection point of the
potential \cite{lw}. The crucial element of both scenarios is the uplifting term $\Delta V=\frac{E}{t^2}$, originating from the $\overline{D3}$-branes,  which is added to the potential in order to break SUSY and cancel cosmological constant in the post-inflationary vacuum. However, in the effective field theoretical
description the $\overline{D3}$-branes break SUSY explicitly.
This is the main drawback of the KKLT stabilization on which racetrack inflation models are based.
In these proceedings we show that both racetrack inflation models can be constructed in a fully supersymmetric framework with the matter field $F$-term as a source of uplifting and SUSY breaking.

It is known that the moduli stabilization at a Minkowski (or dS) minimum can be achieved using $F$-term uplifting \cite{matter}. Nevertheless, successful $F$-term uplifting of inflationary models does not have to be straightforward.
The moduli stabilization is a local problem in a sense that the
only issue which matters is the stability of the potential at
a Minkowski (or dS) stationary point.
On the other hand, the problem of constructing
an inflationary model involves also the global structure of the
potential. The reason is that the Minkowski vacuum and the inflationary
region are in separate domains of the field space. A priori one
cannot be sure that there always exists a trajectory connecting
these two regions. It is especially not obvious that such a
trajectory exists when one increases the dimensionality of the
field space by introducing a matter field. Therefore, it is encouraging that racetrack inflation models can be successfully realized with uplifting from the matter field $F$-term.

\section{Conditions for K\"ahler potential}

It was pointed out in \cite{bo} that the role of uplifting in racetrack inflation models is two-fold. Besides the cancelation of the cosmological constant, uplifting is also crucial for the stability of the vacuum and for fulfilling slow-roll conditions. We explain this point below. The necessary condition for the stable dS vacuum and/or slow-roll inflation depends on the K\"ahler potential in the following way \cite{bo,covi}:
\begin{equation}
\label{etacon}
R(f^i)<\frac{2}{\widehat{G}^2} \,,
\end{equation}
where $R(f^i)\equiv R_{i\ov{j}p\ov{q}}f^if^{\ov{j}}f^pf^{\ov{q}}$
is the sectional curvature of the K\"ahler manifold (defined by the metric given by the second derivative of the
K\"ahler potential $K_{i\ov{j}}$) along the direction of SUSY
breaking and
$f_i\equiv G_i/\widehat{G}$ is the unit vector defining that
direction. We also introduced the quantity
$\widehat{G}\equiv\sqrt{G^iG_i}$ related in a simple way
to the value of the potential: $\widehat{G}^2=3+e^{-G}V$.
For the tree-level K\"ahler potential (\ref{kahler}) the scalar curvature $R_T\equiv R_{T\ov{T}T\ov{T}}f^Tf^{\ov{T}}f^Tf^{\ov{T}}=2/3$ and the necessary condition (\ref{etacon}) is violated for non-negative values of the potential. Nevertheless, racetrack inflation models can be realized because the uplifting term is non-supersymmetric so after adding it to the potential the necessary condition (\ref{etacon}) is no longer valid.

However, our goal is to construct racetrack inflation models in which SUSY is broken spontaneously by the matter field $F$-term and without invoking explicitly SUSY breaking terms. This cannot be achieved if the no-scale K\"ahler potential $K=-3\ln(T+\ov{T}-|\Phi|^2)$ is used because in such a case $R(f^i)=2/3$\ \cite{Grs2} and the necessary condition (\ref{etacon}) is violated.
This fact motivates us to study K\"ahler potentials of the form: $K=K^{(T)}(T,\ov{T})+K^{(\Phi)}(\Phi,\ov{\Phi})$. In such a case the necessary condition (\ref{etacon}) reduces to:
\begin{equation}
\label{thetacon}
R_T\Theta_T^4+R_{\Phi}\Theta_{\Phi}^4<\frac{2}{\widehat{G}^2} \,,
\end{equation}
where $R_i$ are the scalar curvatures of the one dimensional
submanifolds associated with each of the fields and
$\Theta_i^2\equiv G_{i\ov{i}}f^if^{\ov{i}}$
(no summation over $i$ or $\ov{i}$) are the spherical coordinates
parameterizing SUSY breaking.
They satisfy the condition $\Theta_T^2+\Theta_{\Phi}^2=1$.
For the canonically normalized matter field (i.e. with $K^{(\Phi)}=\Phi\ov{\Phi}$), the scalar curvature $R_{\Phi}$ vanishes. Therefore, if the canonically normalized matter field dominates SUSY breaking during inflation (i.e. $\Theta_T^2\ll1$) then the condition (\ref{thetacon}) is satisfied and slow-roll inflation is possible.

\section{Racetrack inflation with Polonyi uplifting}

Consider a racetrack model coupled to the canonically normalized matter field as follows:
\begin{equation}
W=W_0+Ce^{-cT}+De^{-dT}-\mu^2\Phi
\,,
\qquad\qquad
K=-3\ln(T+\overline{T})+\Phi\ov{\Phi}
\,.
\end{equation}
The matter field part of the above model corresponds to the well-known Polonyi model of SUSY breaking. The cancelation of the cosmological constant is due to the fine-tuning of the parameter $\mu$. We found that in the above simple setup both racetrack inflation models can be successfully realized. However, racetrack inflation with Polonyi uplifting is significantly different from original racetrack models \cite{racetrack,lw} with non-SUSY uplifting.

\begin{figure}
\begin{minipage}{0.48\linewidth}
\centering
  \includegraphics[width=\textwidth, height=0.27\textheight]{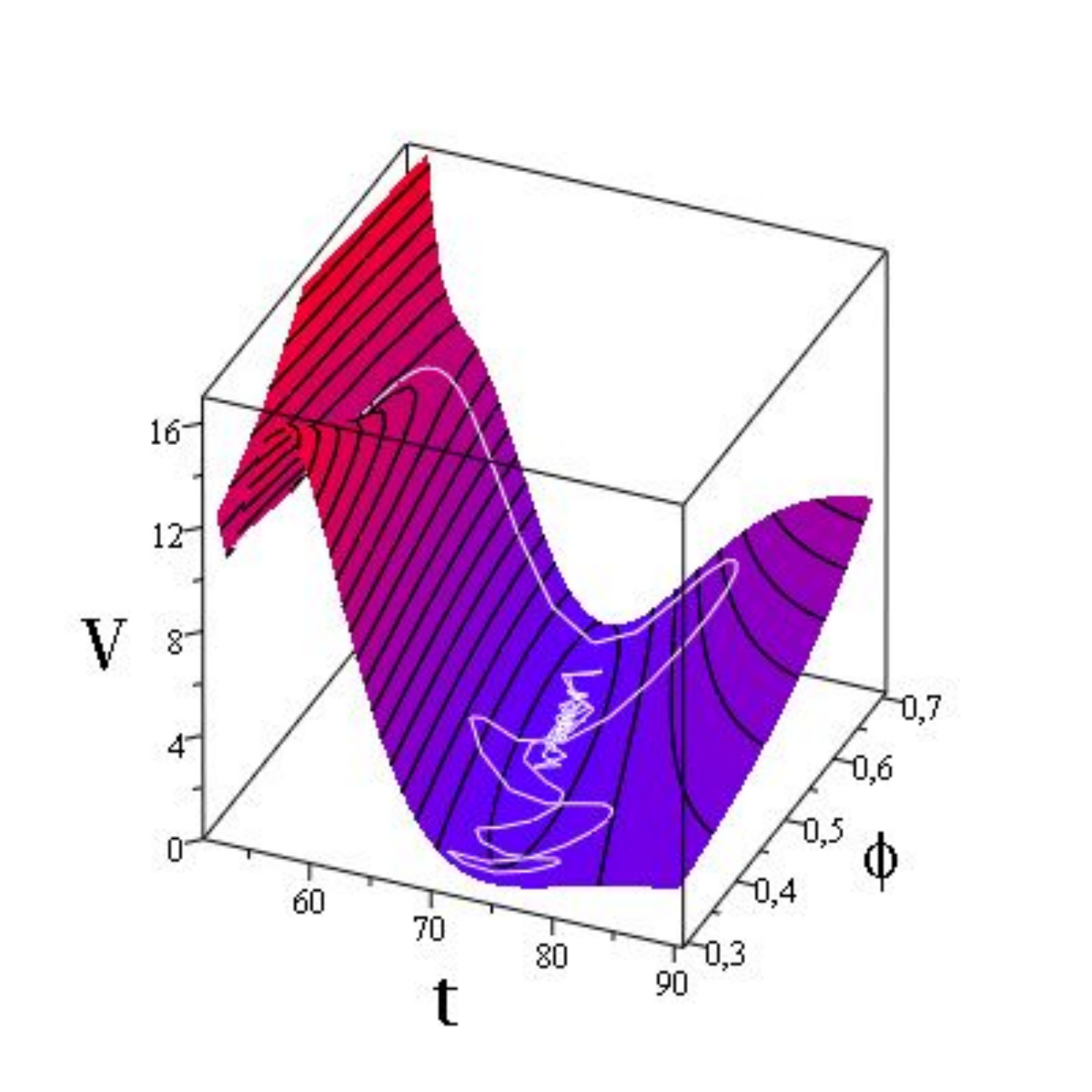}
  \caption{The potential for the inflection
point model coupled to the Polonyi sector for $\tau=\theta=0$. The white curve represents
the field trajectory.
}
\label{KL_polonyi}
\end{minipage}
\hfill
\begin{minipage}{0.48\linewidth}
  \centering
  \includegraphics[width=\textwidth,height=0.27\textheight]{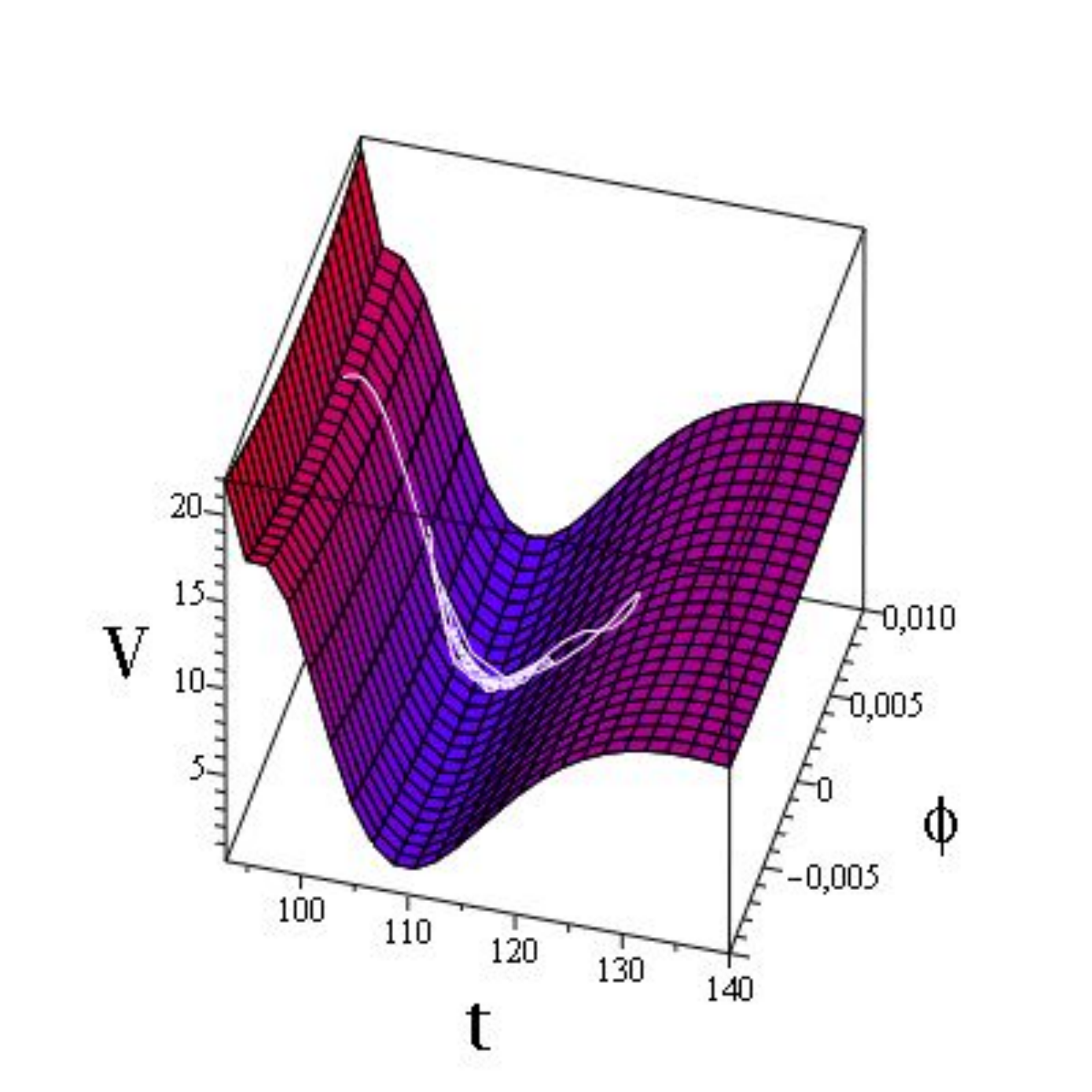}
  \caption{The potential for the inflection point model
coupled to an effective O'Raifeartaigh model for $\tau=\theta=0$. The white curve represents
the field trajectory.
}
\label{KL_okklt}
\end{minipage}
\end{figure}

Let us focus first on the inflection point model. Potential and the trajectory of the inflaton for this model is shown in figure \ref{KL_polonyi}. It can be seen that the real part of the matter field $\phi$ dominates inflation  so the volume modulus is no longer the inflaton. SUSY breaking during inflation is strongly dominated by the matter field $F$-term. In consequence, $R(f^i)\approx0$ and the necessary condition for slow-roll inflation (\ref{etacon}) is easily satisfied.

In the original racetrack inflection point inflation \cite{lw} fine-tuning of parameters, required for obtaining more than $60$ e-folds of inflation, is related to the height of the barrier which prevents the inflaton $t$ from running away to infinity after inflation \cite{bo2}. Avoiding the overshooting problem requires fine-tuning of one parameter (e.g. $W_0$) at the level of $10^{-8}$. In the inflection point model with Polonyi uplifting this problem is less severe and the fine-tuning of $W_0$ at the level of $10^{-3}$ is enough to obtain $60$ e-folds of inflation ending in the Minkowski minimum.

Saddle point racetrack inflation with Polonyi uplifting also significantly differs from the original model \cite{racetrack} with non-SUSY uplifting. In the present case the imaginary part of the matter field $\theta$ is the main component of the inflaton. Fine-tuning of parameters is  of order $10^{-3}$ so it is slightly weaker than in the original model \cite{racetrack} in which fine-tuning is at the level of $10^{-4}$.

\section{Racetrack inflation with O'uplifting}

Let us now consider the following generalization of the previous model:
\begin{equation}
W=W_0+Ce^{-cT}+De^{-dT}-\mu^2\Phi
\,,
\qquad\qquad
K=-3\ln(T+\overline{T})+\Phi\ov{\Phi}-\frac{(\Phi\ov{\Phi})^2}{\Lambda^2}
\,.
\end{equation}
The model with Polonyi uplifting is recovered in the limit $\Lambda\to\infty$. The matter field sector of the above model can be treated as an effective quantum corrected O'Raifeartaigh model with the superpotential $W^{(O')}=mXY+\lambda\Phi X^2-\mu^2\Phi$ in which the heavy fields $X$ and $Y$ have been integrated out \cite{okklt}. The parameter $\Lambda$ corresponds to the mass scale of the fields that have been integrated out so a natural value of $\Lambda$ is much smaller than one (in Planck units). The parameter $\mu$ is, again, adjusted in such a way that the cosmological constant at the post-inflationary vacuum (almost) vanishes. The value of the real part of the matter field during inflation, as well as at the Minkowski minimum, is $\phi\sim{\mathcal O}(\Lambda^2)$. So, in the region important for inflation the following hierarchy is present: $\phi\ll\Lambda\ll1$.
One can show that in the limit $\phi\ll\Lambda\ll1$ the mass matrix is nearly diagonal and the matter field is heavier than the volume modulus. In consequence, the matter field is approximately decoupled from the inflationary dynamics.

Both racetrack inflation models resemble original ones \cite{racetrack,lw} with non-SUSY uplifting. As an example illustrating this fact we present the plot for the inflection point model in figure \ref{KL_okklt}. It can be seen that the volume modulus $t$ plays the role of the inflaton (as in the original model \cite{lw}) while the matter field is almost frozen during inflation. Another similarity to the original model is  that the fine-tuning of parameters is related to the height of the barrier which separates the Minkowski vacuum from the runaway region. It is worth to note that in the inflection point model with O'uplifting (or with uplifting from $\ov{D3}$-brane) it is possible to arrange inflation with arbitrary low scale. However, this would require extremely large values of parameters $C$ and $D$.
The saddle point model with O'uplifting is also similar to the corresponding one with non-SUSY uplifting. In particular, axion $\tau$ dominates inflation and fine-tuning of $W_0$ is of order $10^{-4}$.

\section{Conclusions}

In these proceedings we have shown that both racetrack inflation models \cite{racetrack,lw} can be constructed in a fully supersymmetric framework with the matter field $F$-term being a source of uplifting and SUSY breaking. The details of inflationary scenarios depend on the choice of the matter field sector. If the Polonyi model is chosen for the uplifting sector, the real (imaginary) part of the matter field dominates the inflection (saddle) point racetrack inflation. With this kind of uplifting the fine-tuning of parameters is significantly weaker than in models with non-SUSY uplifting (especially in the inflection point model). On the other hand, if the O'Raifeartaigh model is responsible for uplifting, the matter field is decoupled from the inflationary dynamics and racetrack inflation models are similar to the original ones but with one important difference: SUSY is now broken spontaneously. In these models, the volume modulus is the inflaton even though SUSY breaking is dominated by the matter field $F$-term.
More detailed analysis of models presented here can be found in \cite{bo3}.

\section*{Acknowledgments}
M.B. would like to thank M.~Olechowski for many fruitful discussions and collaboration. M.B was partially supported by the EC 6th Framework Project MRTN-CT-2006-035863 ``The Origin of Our Universe: Seeking Links between Fundamental Physics and Cosmology'' and by Polish MNiSzW scientific research grants N N202 103838 and N N202 285038.

\end{document}